\documentclass[fleqn,usenatbib]{mnras}

\usepackage{times}

\usepackage[T1]{fontenc}
\usepackage{ae,aecompl}

\usepackage{graphicx}	
\usepackage{amsmath}	
\usepackage{amssymb}	
\usepackage{aas_macros}

\newcommand{\source}{J1753.5}
\newcommand{\swift}{\textit{Swift}}

\title[Optical Spectroscopy of Swift J1753.5-0127]{No evidence for a low-mass black hole in Swift J1753.5-0127}

\author[A. W. Shaw et al.]{
A. W. Shaw,$^{1}$\thanks{E-mail: A.Shaw@soton.ac.uk}
P. A. Charles,$^{1}$
J. Casares,$^{2,3,4}$
and J. V. Hern\'{a}ndez Santisteban$^{1}$
\\
$^{1}$Department of Physics and Astronomy, University of Southampton, Southampton SO17 1BJ, UK\\
$^{2}$Instituto de Astrof\'{i}sica de Canarias, E-38205 La Laguna, Tenerife, Spain\\
$^{3}$Departamento de astrof\'{i}sica, Univ. de La Laguna, E-38206 La Laguna, Tenerife, Spain\\
$^{4}$Department of Physics, Astrophysics, University of Oxford, Denys Wilkinson Building, Keble Road, Oxford OX1 3RH, UK
}

\date{Accepted XXX. Received YYY; in original form ZZZ}

\pubyear{2016}

\begin{document}
\label{firstpage}
\pagerange{\pageref{firstpage}--\pageref{lastpage}}
\maketitle

\begin{abstract}
We present high-resolution, time-resolved optical spectroscopy of the black hole X-ray transient Swift J1753.5-0127. Our optical spectra do not show features that we can associate with the companion star. However we do observe broad, double-peaked emission lines, typical of an accretion disc. We show that the mass of the compact object is likely $>7.4\pm1.2M_{\odot}$, much higher than previous suggestions of a low-mass ($<5M_{\odot}$) black hole. 
\end{abstract}

\begin{keywords}
black hole physics -- X-rays: binaries --  X-rays: individual: Swift J1753.5-0127
\end{keywords}



\section{Introduction}
Galactic black hole X-ray transients (BHXRTs) are low-mass X-ray binaries (LMXBs) in which a black hole (BH) accretes material from a donor star via an accretion disc. Approximately 75\% of all LMXB transients are believed to harbour a BH \citep{McClintock-2006} and are characterised by long periods of quiescence (years to decades) followed by X-ray outbursts which can increase the luminosity by several orders of magnitude. BHXRTs have proven to be important in studying LMXBs, as in quiescence they provide the opportunity to study the donor itself, which is mostly impossible in luminous, persistent XRBs \citep{Charles-2006}.\\
\indent Using this technique of studying largely quiescent BHXRTs, and adding in the mass determinations in high mass X-ray binaries (HMXBs) there are now more than $\sim20$ BH mass determinations (e.g. \citealt{Casares-2014}). The distribution of these masses has become the subject of intense scrutiny from both an observational and theoretical viewpoint as, when combined with the substantial number of accurate neutron star mass measurements, there appears to be a dearth of compact objects with masses in the range $2-5M_{\odot}$ (\citealt{Ozel-2010,Farr-2011} and references therein). This has even been referred to as the ``mass gap", and has prompted theoretical explanations that derive from the nature of the supernova mechanism that produced the BHs, ranging from convective instabilities at the time of the explosion to `failed' supernovae in a certain mass range of red supergiants \citep{Belczynski-2012,Ugliano-2012,Kochanek-2014}. Consequently, there is considerable interest in either (a) accounting for this mass gap as a selection bias of some form (e.g. \citealt{Farr-2011,Kreidberg-2012}), or (b) finding objects with masses in or close to this range. Hitherto, the lowest mass BHXRT has been considered to be GRO J0422+32 ($\sim5M_{\odot}$), but there has been a significant uncertainty in this value as a result of its poorly determined inclination (see \citealt{Casares-2014} and references therein). The search has therefore been on for a low mass BHXRT with a high binary inclination, and hence a more accurate mass determination. \\
\indent The subject of this work, Swift J1753.5-0127 (hereafter \source) was discovered by the \swift~ Burst Alert Telescope (BAT; \citealt{Barthelmy-2005}) in 2005 \citep{Palmer-2005} as a hard-spectrum ($\gamma$-ray source) transient at a relatively high Galactic latitude (+12$^{\circ}$ ). The source luminosity peaked within a week, at a flux of $\sim$200 mCrab, as observed by the \textit{Rossi X-Ray Timing Explorer} (\textit{RXTE}) All Sky Monitor (ASM; 2--12 keV) \citep{CadolleBel-2007}. The source was also detected in the UV, with \swift's Ultraviolet/Optical Telescope (UVOT; \citealt{Still-2005}), and in the radio with MERLIN \citep{Fender-2005}. A $R\sim15.8$mag optical counterpart was identified by \citet{Halpern-2005}, who noted that it had brightened by at least 5 magnitudes (as it is not visible in the Digitized Sky Survey; DSS), thereby establishing \source~as an LMXB.\\
\indent Subsequent time-resolved photometry of the optical counterpart \citep{Zurita-2008} revealed $R$-band modulations on a period of 3.24h, which were interpreted as a superhump period, $P_{\mathrm{sh}}$. Such periodicities are seen in high mass ratio (compact object/donor) CVs (the SU UMa systems) where it has been found by \citet{Patterson-2005} that the period differential (defined as $\epsilon = (P_{\rm sh}-P_{\rm orb})/P_{\rm orb}$) is a function of the mass ratio.  For the BH LMXBs $\epsilon$ is very small, and so the superhump period is $<$2\% longer than the system $P_{\rm orb}$. The presence of superhumps in the optical light curves of \source~are indicative of an accretion disc that is precessing due to perturbations by the companion and have been seen to occur in other BHXRTs such as XTE J1118+480 \citep{Zurita-2002}.\\
\indent More importantly for our study, the $R$-band modulations suggested that \source~was a high inclination system (although not actually eclipsing or dipping) and hence held the potential for an accurate mass determination. This could be accomplished either (a) once it had returned to quiescence (although this would be challenging given its absence on the DSS), or (b) during outburst if it displayed fluorescence emission features on the surface of the donor whilst still X-ray active (e.g. \citealt{Cornelisse-2008}).\\
\indent Almost immediately after its peak the X-ray flux of \source~started declining, but it then remained roughly constant at $\sim20$ mCrab (2--12 keV) for over 6 months rather than returning to quiescence as might have been expected for a typical BHXRT \citep{McClintock-2006}. The source has still not returned to quiescence $\sim11$ years after its initial discovery, and has instead exhibited significant long-term ($>400$d) variability over the course of its prolonged `outburst' \citep{Shaw-2013a}.\\
\source~has remained as a persistent LMXB in a hard accretion state for the majority of this time, however it has experienced a number of short-term spectral softenings, characterised by an increase in the temperature of the inner accretion disc and simultaneous steepening of the power-law component in the X-ray spectrum \citep{Yoshikawa-2015}. Investigation of the source during one such event with \textit{RXTE} revealed that it had transitioned to a hard intermediate accretion state. However, unlike the majority of BHXRTs, Swift J1753.5-0127 did not continue towards an accretion disc dominated soft state and instead returned to the hard state \citep{Soleri-2013}. In early 2015, the source appeared to undergo another state transition when the \textit{Swift}-BAT flux appeared to drop to its lowest levels since the source's discovery \citep{Onodera-2015}. Subsequent follow-up with the \textit{Swift} X-ray Telescope (XRT; \citealt{Burrows-2005}), \textit{XMM-Newton} \citep{Jansen-2001} and the \textit{Nuclear Spectroscopic Telescope Array} (\textit{NuSTAR}; \citealt{Harrison-2013}) revealed that \source~had transitioned to one of the lowest luminosity soft states recorded in LMXBs \citep{Shaw-2015a,Shaw-2016b}.\\
\indent With a large ($\Delta R\sim5$ mag) optical increase at outburst, we would not expect to detect any spectroscopic signatures of the donor whilst it was active, due to the optical light being dominated by the accretion disc. \citet{Durant-2009} confirmed this with spectroscopic observations revealing a smooth optical continuum and no evidence for features that could be associated with the donor. With no detectable fluorescence emission either, it had therefore not been possible to obtain any direct evidence of the compact object mass. However, INTEGRAL observations highlighted the presence of a hard power-law tail up to $\sim600$ keV, very typical of a black hole candidate (BHC) in the hard state \citep{CadolleBel-2007}. Also, the power density spectrum from a pointed \textit{RXTE} observation revealed a 0.6 Hz quasi-periodic oscillation (QPO) with characteristics typical of BHCs \citep{Morgan-2005}. QPOs have also been seen at 0.08 Hz in optical data \citep{Durant-2009} as well as in a number of X-ray observations after the initial outburst had declined \citep{Ramadevi-2007,CadolleBel-2007}.\\
\indent Remarkably, given the above summary of BHXRT properties, \citet[][hereafter N14]{Neustroev-2014} have presented evidence that \source~ does contain a low mass ($<5M_{\odot}$) BH, based on their discovery of narrow optical features (in both emission and absorption) which they associate with the donor, despite such features not being identifiable or visible in previous spectroscopic studies \citep{Durant-2009}. Given the considerable potential importance of the identification of a high inclination, low-mass BH in the ``mass gap", we therefore undertook a spectroscopic study of \source, using significantly higher spectral resolution so as to investigate its properties in much greater detail, and at the very least attempt to confirm the low-mass BH candidacy of \source. We therefore focussed on the donor's features reported by N14, and report here our inability to reproduce any of their results.

\section{Observations and Analysis}
\source~was observed from 2015 June 14 21:39:36 UT to June 15  04:44:55 UT (MJD 57187.903--57188.198) with the Intermediate dispersion Spectroscopic and Imaging System (ISIS) on the 4.2-m William Herschel Telescope (WHT) at Observatorio del Roque de los Muchachos, La Palma, Spain. We obtained 25$\times$900s exposures covering a total spectral range 4174--7134\AA, utilising the R600B and R600R gratings and a 1" slit in photometric conditions of good ($\sim1$") seeing. We used \scriptsize IRAF \normalsize \citep{IRAF} to perform standard reduction techniques to achieve wavelength calibration, cosmic ray removal using the external task ``lacos" \citep{van_Dokkum-2001} and extraction of the 1-dimensional spectra. CuAr+CuNe comparision arcs were obtained to calibrate the wavelength scale, achieving a central dispersion of  0.43 and 0.49 \AA~pixel$^{-1}$ in the blue and red arms, respectively. We obtain a spectral resolution of 1.55 and 1.66\AA~(FWHM) in the blue and red arms, respectively. We also observed the spectral type standard stars HR 4949 (M5III), HR 4986 (M0III), HR 4929 (K0III) and HR 4962 (K5III), whose spectra were reduced and extracted in the same way. We flux calibrated the averaged spectrum using the nearby flux standard star BD +33 2642 \citep{Oke-1990}. We note that the observations were performed when \source~was in the soft state \citep{Shaw-2016b}.\\
\indent Much of the analysis of the extracted spectra was performed using Tom Marsh's \textsc{molly} software package.\footnote{http://www2.warwick.ac.uk/fac/sci/physics/research/astro/people/marsh/software/} Computation of cross-correlation was performed separately for each arm of the spectrograph using the task \scriptsize XCOR\normalsize, which computes the velocity shift of the target spectrum with respect to a template spectrum. We created 25 template spectra, each spectrum created by averaging all of the spectra of \source~except for the target spectrum used for the calculation of the cross-correlation. For example, we cross-correlated spectrum 1 with a template created from the average of spectra 2--25. This removes the possibility of a false positive correlation. We cut the spectra in the blue arm above 5500\AA~due to large-scale variations in the flat-fields. The individual spectra and the template spectra were rebinned onto a uniform velocity scale in each arm and the continuum was subtracted by fitting a low-order spline to the data. Spectral features which may have affected the calculation of the cross-correlation function such as interstellar lines, H$_{\alpha}$ and He\small II \normalsize4686\AA~were masked out of the spectra and the templates before the calculation was performed. In the blue arm, cross-correlation was performed between 4174--5500\AA~and in the red arm between 5618--7134\AA.\\
\indent We also performed Doppler Tomography using the \textsc{python} implementation of Tom Marsh's \textsc{doppler} package.\footnote{https://github.com/trmrsh/trm-doppler} We used the continuum subtracted spectra from each arm, trimmed around the lines of interest to compute the maps. To create the reconstructed trailed spectra from the resultant maps, we used the \textsc{doppler} task \textsc{comdat}.

\section{Results}

\begin{figure*}
\includegraphics[scale=1.0,trim={1cm 3cm 0 3cm}]{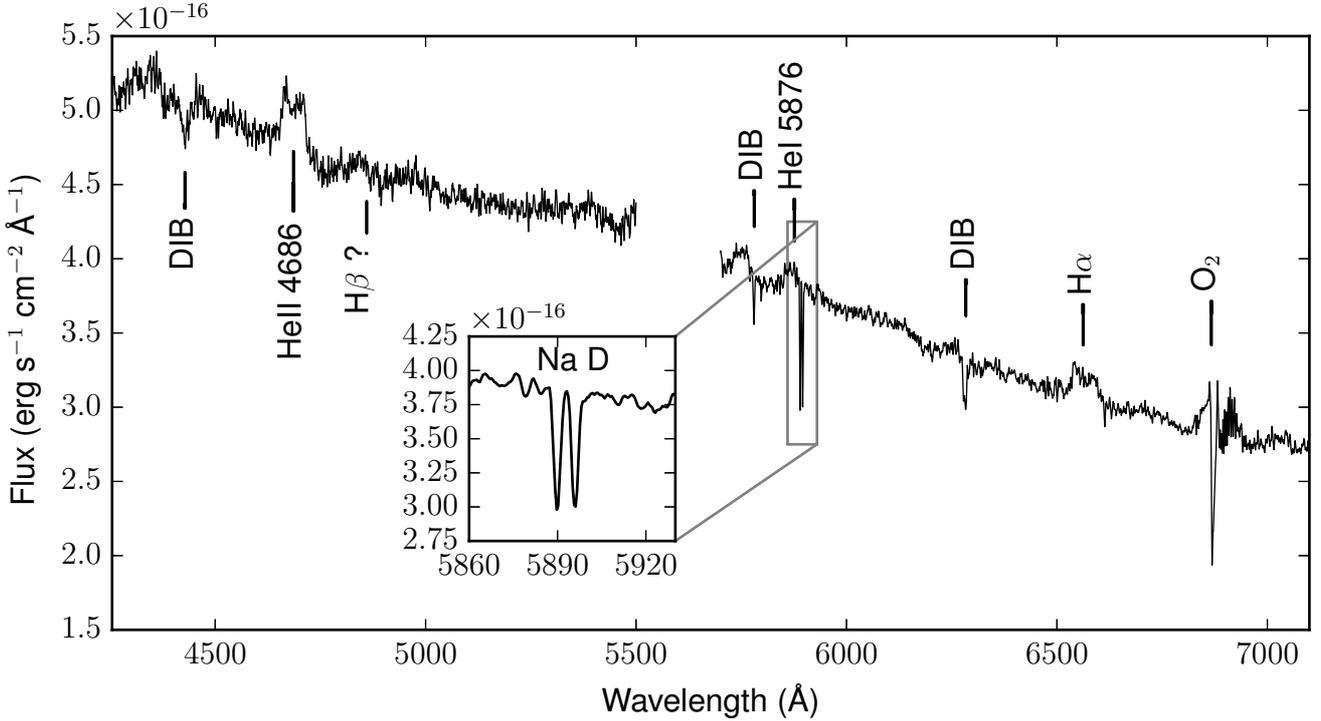}
\caption{Flux calibrated averaged optical spectrum of \source. Notable lines have been annotated. The inset shows a zoom of the Na D interstellar lines, highlighting that the two lines are resolved.}
\label{spec}
\end{figure*}

\begin{figure}
\includegraphics[scale=0.45,trim={0.75cm 0 0 0}]{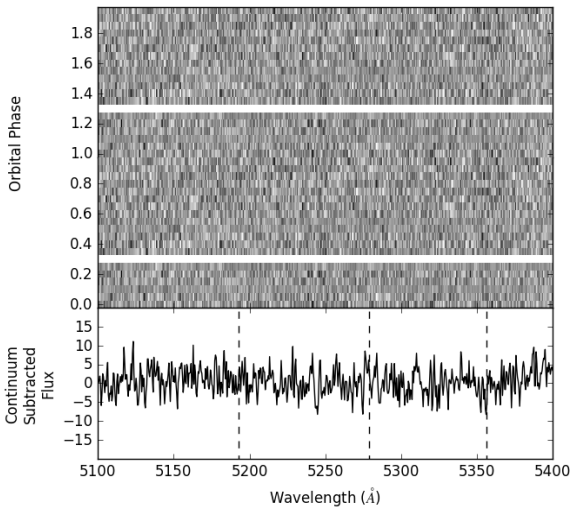}
\caption{Top: The trailed spectrum of \source~ in the 5100-5400\AA~range, phase-folded on the $P_{\mathrm{orb}}$ determined by N14. White indicates emission. Two cycles are shown for clarity. Bottom: The continuum subtracted, averaged spectrum of \source. The dashed lines highlight the location of the narrow absorption and emission features seen by N14, which they associated with the secondary star.}
\label{trail}
\end{figure}

The flux-calibrated spectrum of \source~is presented in Fig. \ref{spec} and exhibits strong double peaked He\small II \normalsize4686\AA~and H$_{\alpha}$ emission lines along with evidence of weak H$\beta$ emission and He \textsc{i} 5876\AA. A number of diffuse interstellar bands (DIBs) are also present in the spectrum, most notably at 4428\AA~and 6283\AA~as well as the (resolved) interstellar Na D lines. We measure the equivalent widths (EWs) of the Na D interstellar lines to be $0.58\pm0.03$ and $0.55\pm0.02$ \AA~for Na D$_1$ and D$_2$ at 5889 and 5895\AA, respectively. This represents a decrease in EW from previous measurements \citep{Durant-2009} and such variability is indicative of some sodium being intrinsic to the system. The Fraunhofer B absorption features due to O$_2$ are also apparent at 6867\AA. We find no absorption features in the averaged spectrum which might be identified with the secondary star.\\
\subsection{Companion star features}
N14 presented evidence of `unidentified' narrow absorption and emission features in the spectrum of \source, which showed a sinusoidal modulation over their claimed 2.85h $P_{\mathrm{orb}}$. They concluded that the features, the strongest of which were observed at 5193, 5279 and 5356.1\AA, were associated with the companion star and therefore the observed modulation was due to its orbital motion.\\
\indent In order to investigate this claim, we phase-folded our spectra on the $P_{\mathrm{orb}}$ and ephemeris calculated by N14. We created a trailed spectrum of the region where such features were noted by N14 (5100-5400\AA), and present this in Fig. \ref{trail}. Studying the trails in Fig. \ref{trail}, we see no evidence for any of the moving features seen by N14 in our WHT spectra. We also note that there is also no evidence for such features when we phase-fold the data on the $P_{\mathrm{orb}}$ determined by \citet{Zurita-2008}. To further examine the spectra for the presence of the features we also attempted to fit a number of gaussians centred on the wavelengths of the observed lines to each spectrum.This was a repeat of the analysis of N14, but found that the fits did not converge with sensible results, confirming our non-detection of these features.

\subsection{Cross-correlation}
\begin{figure*}
\centering
\begin{minipage}[c]{0.5\textwidth}
\centering
\includegraphics[scale=0.45]{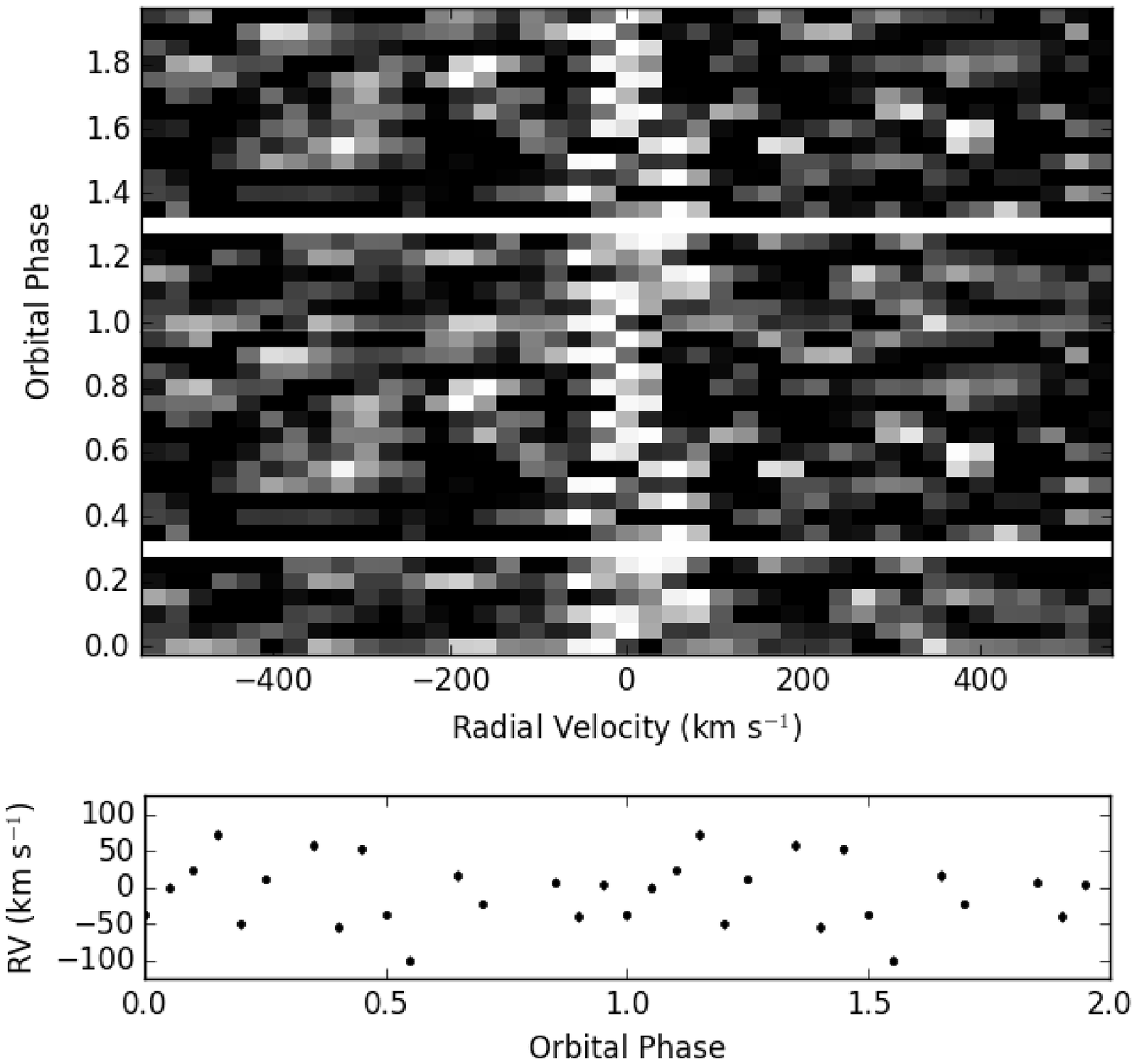}
\end{minipage}%
\begin{minipage}[c]{0.5\textwidth}
\centering
\includegraphics[scale=0.45]{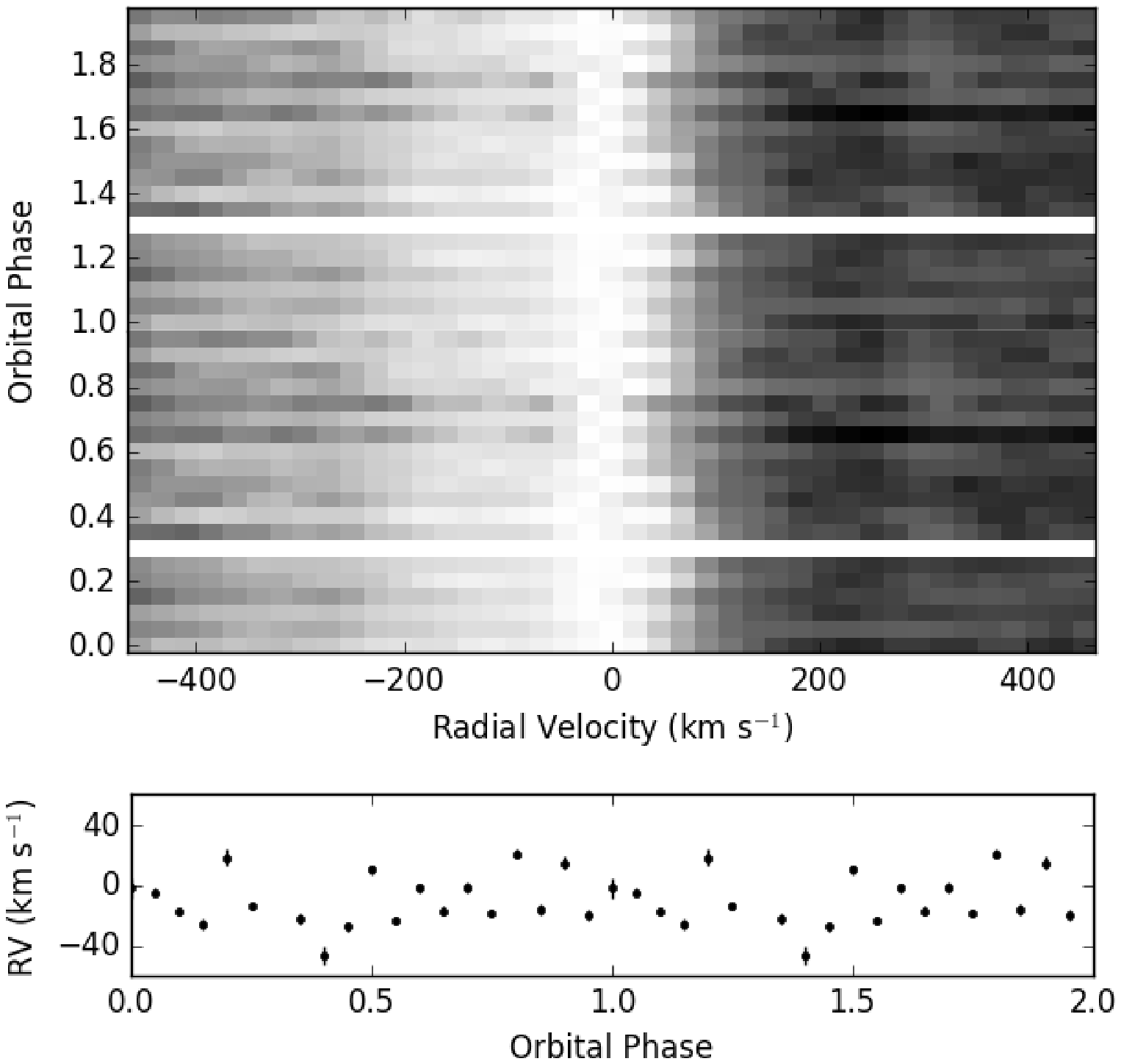} 
\end{minipage}
\caption{Left: Trailed, phase-folded CCF spectra from the blue arm. Right: Trailed, phase-folded CCF spectra from the red arm. Both datasets have been folded on the $P_{\mathrm{orb}}$ determined by N14. The lower panels of each trailed spectrum show the peak radial velocity obtained from the CCFs for each phase. Two phases are shown for clarity.}
\label{CCF}
\end{figure*}

Cross-correlation was performed using the \scriptsize MOLLY \normalsize task \scriptsize XCOR \normalsize as detailed above. The resultant cross-correlation function (CCF) spectra were then phase-folded on the $P_{\mathrm{orb}}$ and ephemeris of N14, for comparison. We created trailed CCF spectra for each arm, which are presented in Fig. \ref{CCF}. The trailed spectra show no large deviations from 0 km s$^{-1}$ and the sinusoidal modulation as seen by N14 is absent. We also note that there is no obvious modulation when the data are folded on $P_{\mathrm{orb}}=3.24$h \citep{Zurita-2008}.\\
\indent We also cross-correlated the spectra of \source~with four spectral type standards observed on the same night. We find no correlation with any of the spectral type templates, indicating that there are likely no features present in the spectrum of \source~that can be associated with the range of late-type stars (K and M-type) most likely to be donors in this system.\\

\subsection{Disc emission lines}
The optical spectrum exhibits He\small II \normalsize4686\AA~and H$_{\alpha}$ emission lines. The lines are very broad and both exhibit a double-peaked structure, very typical of emission features originating in an accretion disc \citep{Smak-1981,Horne-1986a}, however they are weak (EW $\sim4$\AA~for a double gaussian fit to each line). To examine the line profiles, first we fit a double gaussian to the emission lines in the phase-averaged spectrum, adopting a non-linear least-square approach. From this we can obtain an estimate of the rotational velocity at the outer rim of the disc, $v_D\sin i$, by measuring the observed peak-to-peak separation \citep{Smak-1981,Warner-1995}. Taking H$_{\alpha}$, as this originates in the outermost regions of the accretion disc, we estimate $v_D\sin i=865\pm25$km s$^{-1}$. For comparison, from the double-peaked He\small II \normalsize4686\AA~emission we calculate $v_D\sin i=1115\pm29$km s$^{-1}$, as expected since He \textsc{ii} is formed closer in to the compact object.\\
\indent The double-peak separation of the H$_{\alpha}$ line profile can be used to estimate the radial velocity semi-amplitude, $K_2$ of the companion star in BHXRTs \citep{Orosz-1994,Orosz-1995}. In quiescent BHXRTs, the ratio $v_D/K_2$ has been shown to be $\simeq1.1-1.25$. For \source, we therefore find a very conservative lower limit of $K_2=692\pm20$ km s$^{-1}$. It must be noted that this value is likely underestimated due to the source not yet being in quiescence. In outburst, the accretion disc expands to a radius larger than that in quiescence, therefore $v_D$ is smaller in outburst. $K_2$ can be used to constrain the mass of the compact object, $M_1$, by calculating the mass function:
\begin{equation}
f(M_1)=\frac{P_{\mathrm{orb}}K_2^3}{2\pi G}=\frac{M_1^3\sin^3i}{(M_1+M_2)^2}
\label{massfunc}
\end{equation}
where $i$ is the inclination of the system and $M_2$ is the mass of the companion. In the case of \source, which is a LMXB, then the mass function represents a firm lower limit to $M_1$. Using $K_2=692\pm20$ km s$^{-1}$ and $P_{\mathrm{orb}}=3.2443\pm0.0010$h (actually the superhump period only slightly larger than $P_{\mathrm{orb}}$; \citealt{Zurita-2008}); \citealt{Zurita-2008}), we obtain $M_1>4.6\pm0.3M_{\odot}$. This again strongly indicates that the compact object in \source~is a BH. The fact that this is a very conservative lower limit on the mass reinforces the case for a BH primary. \\
\indent Recently, \citet{Casares-2015} discovered that the FWHM of a single gaussian fitted to the H$_{\alpha}$ line profile is tightly correlated with $K_2=0.233(13)\mathrm{FWHM}$ in quiescent BHXRTs. As above, this relation is only valid for quiescence, but as the accretion disc expands during an outburst, the FWHM gets smaller, thus FWHM$_{\mathrm{(outburst)}}$ < FWHM$_{\mathrm{(quiescence)}}$. We find $\mathrm{FWHM}=3470\pm214$ km s$^{-1}$, which translates to $K_2>808\pm67$ km s$^{-1}$. When combined with $P_{\mathrm{orb}}$ we find $M_1>7.4\pm1.2M_{\odot}$. This again gives us a conservative lower limit on the mass of the compact object, higher than that calculated from the double peak separation. \\
\indent Another way of constraining the mass of the compact object in \source~is by scaling the FWHM of the $H_{\alpha}$ line with that of another system with well determined system parameters, such as XTE J1118+480. Since FWHM scales with $\sin i \left (\frac{M_1}{P_{\mathrm{orb}}}\right)^{1/3}$ then:

\begin{multline}
M_1\sin^3i_{(\mathrm{J1753})} = M_1\sin^3i_{(\mathrm{J1118})}\left(\frac{\mathrm{FWHM_{(J1753)}}}{\mathrm{FWHM_{(J1118)}}}\right)^3 \\
\times\left(\frac{P_{\mathrm{orb (J1753)}}}{P_{\mathrm{orb (J1118)}}} \right)
\end{multline}

Adopting for XTE J1118+480 the following parameters: $M_1 = 7.5M_{\odot}$, $i=73^{\circ}$ \citep{Khargharia-2013}, $P_{\mathrm{orb}}=0.170$d \citep{Gonzalez-Hernandez-2012} and FWHM=2850 km s$^{-1}$ \citep{Casares-2015}. Combining these values with our parameters for \source, the scaling yields $M_1\sin^3i_{(\mathrm{J1753})}\approx9.4M_{\odot}$ and even $\approx8.3M_{\odot}$ if $P_{\mathrm{orb}}=2.85$h is assumed. Again, these values are lower limits because the FWHM has been measured in outburst and hence is an underestimate of the quiescent value. Therefore, the scaling of FWHM with XTE J1118+480 suggests that, in contrast with N14, the BH in \source~is rather massive, even in the extreme case of an edge-on system.\\
\indent We can also use the disc emission lines to investigate the motion of the compact object to the companion, i.e. $K_1$. To do this we cross-correlated the individual spectra with the phase-averaged spectrum, masking all but the He \textsc{ii} and H$_{\alpha}$ lines. However, the resultant CCFs were consistent with $0$ km s$^{-1}$ within $1\sigma$ statistical uncertainties of $22$ km s$^{-1}$, indicating that we could not detect the orbital motion of the BH. This is contrary to N14, who claimed a value of $K_1=52\pm10$ km s$^{-1}$ by measuring the centroid of He \textsc{ii} 4686\AA~emission feature, although N14 cautioned that the parameters obtained are plagued with systematic errors. We conclude that, though our spectra are higher resolution than those of N14, we cannot detect the orbital motion of the BH as we are limited by the signal-to-noise ratio of the individual spectra. Furthermore, the the overall emission line profile may not accurately trace the motion of the central star \citep[see eg.][]{Orosz-1994}.

\subsubsection{Doppler Tomography}
\begin{figure*}
\centering
\includegraphics[scale=0.35,trim={5cm 0 4cm 0}]{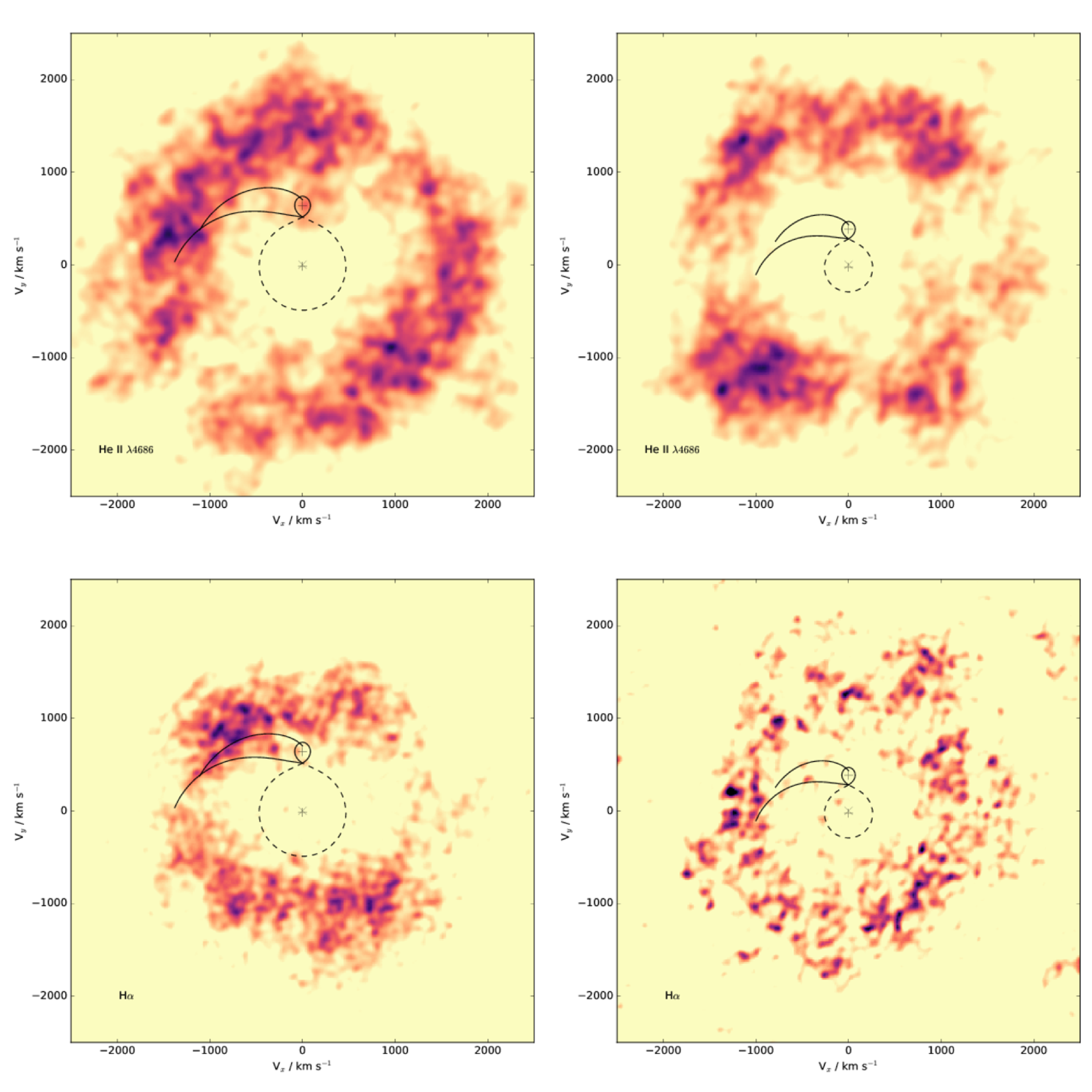}
\caption{Doppler tomograms for the He \textsc{ii} 4686\AA~(upper) and H$_{\alpha}$ (lower) emission lines. The two left panels were constructed utilising the system parameters of \citet{Zurita-2008}, the two right panels were constructed using those of N14.}
\label{tomograms}
\end{figure*}

\begin{figure}
\centering
\includegraphics[scale=0.17]{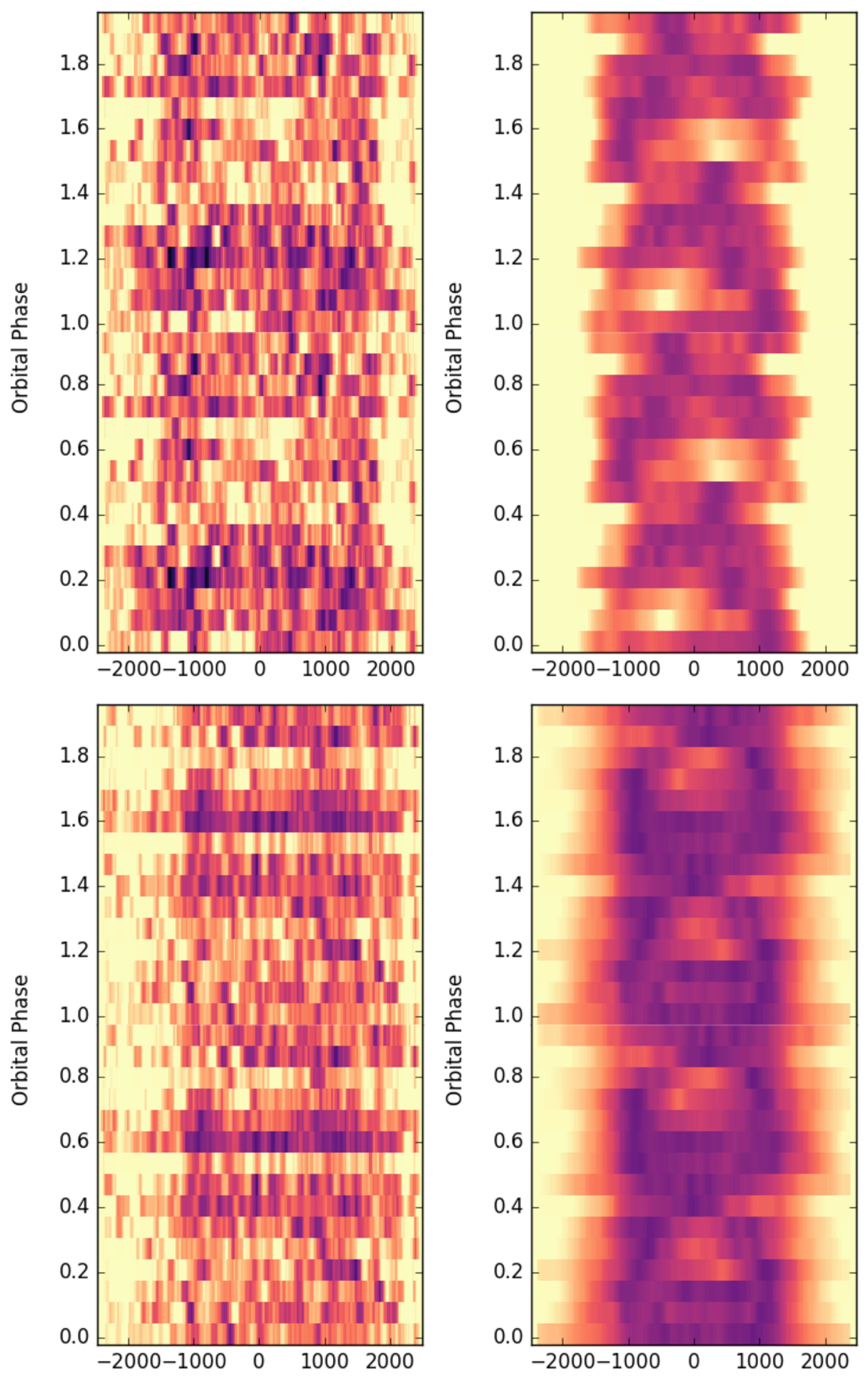}
\caption{Observed and reconstructed trailed spectra for the He \textsc{ii} 4686\AA~(upper) and H$_{\alpha}$ (lower) emission lines. The trails were constructed using the system parameters of \citet{Zurita-2008}.}
\label{reconstructed}
\end{figure}

We can study the line profiles in more detail by examining the trailed spectra, which show variations of the blue and red peaks of the double-peaked line profiles. The orbital variation of the lines enables us to study the disc structure using the technique of Doppler tomography. We utilised the maximum entropy implementation of Doppler tomography using Tom Marsh's \textsc{doppler} software package (see \citealt{Marsh-1988} for technical details). The resultant tomograms are presented in Fig. \ref{tomograms}, where we have marked the Roche lobe of the compact object (dashed line) and the secondary (solid line) using the system parameters of \citet{Zurita-2008}. The system parameters are not well known for \source, so the plotted Roche lobes are estimates, nevertheless, they represent typical values for LMXB BH systems (see \citealt{Casares-2014}), which we believe to be more appropriate for \source~(see section \ref{sec:mass}). The associated observed and reconstructed trailed spectra are presented in Fig. \ref{reconstructed}.\\
\indent The tomograms show clearly the disc structure where the double peaked emission line profiles originate. The radii of the annuli are different for He \textsc{ii} and H$_{\alpha}$, with H$_{\alpha}$ exhibiting smaller velocities - indicative of the smaller peak-to-peak separation in the line. The emission for both lines is roughly symmetric, showing none of the enhanced structure seen by N14. We constructed tomograms using the system parameters calculated by N14, also presented in Fig. \ref{tomograms}. However, they do not exhibit the same `clean' disc structure, instead showing four distinct regions of enhanced `clumpy' emission rather than the uniform disc we see using the parameters from \citet{Zurita-2008}. \\
\indent It is important to note that the tomograms computed with the system parameters of N14 in Fig. \ref{tomograms} show a disc constrained to emit only from an inner fraction of its Roche lobe. With ongoing mass transfer (for this very extended outburst) it is far more likely that emission extends out to velocities associated with the orbit of the donor. Such structure is then seen in the tomograms calculated using the parameters of \citet{Zurita-2008} as is also clear in those of XTE J1118+480 (see Fig. 5 of \citealt{Torres-2002}).

\section{Discussion}
\subsection{Orbital Period}
\begin{figure}
\centering
\includegraphics[scale=0.4]{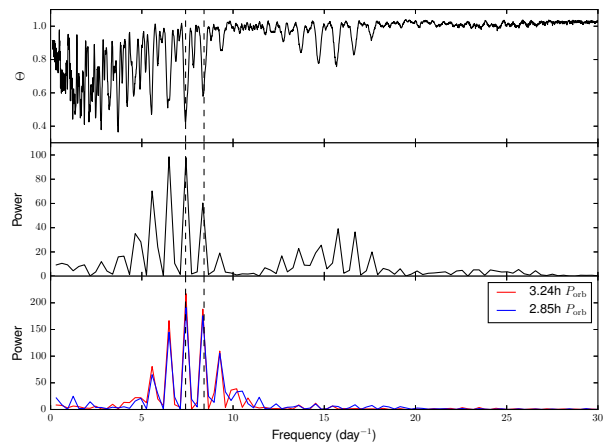}
\caption{\emph{Top}: PDM periodogram of the $R$-band nightly photometry from \citet{Zurita-2008} which has been filtered to match the duration of the spectroscopic observations of N14 (see text for full description). \emph{Centre}: Lomb-Scargle periodogram of the same photometry. \emph{Bottom}: Lomb-Scargle periodograms of light curves created by simulating a sine curve with a period of 3.24h (Red) and 2.85h (Blue) using the same sampling as the filtered photometry in the upper two panels. The vertical dashed lines represent the two determined values of $P_{\mathrm{orb}}$; 3.24h (left) and 2.85h (right). }
\label{periodograms}
\end{figure}

\begin{figure}
\centering
\includegraphics[scale=0.4]{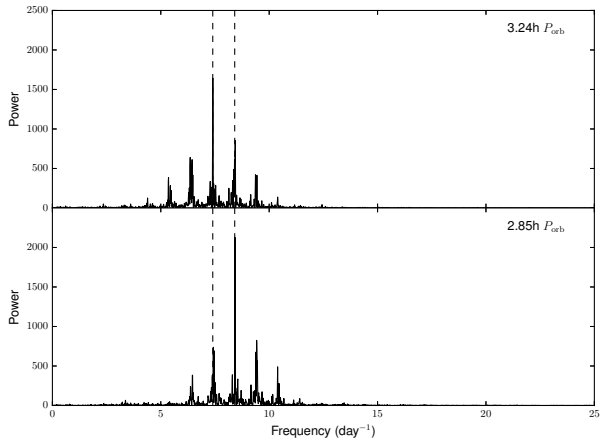}
\caption{Lomb-Scargle periodogram of light curves created by simulating a sine curve with a period of 3.24h (\emph{Top}) and 2.85h (\emph{Bottom}) using the same sampling as the light curves presented by \citet{Zurita-2008}. The vertical dashed lines represent the two proposed values of $P_{\mathrm{orb}}$; 3.24h and 2.85h at frequencies of $\sim7.4$ and $\sim8.4$ day$^{-1}$.}
\label{simulated_LS}
\end{figure}

N14 determined a period of $2.85$h by performing time-series analysis on both their spectroscopic and photometric data. The source is known to show strong modulations in its optical light curve \citep{Zurita-2008}, but N14 also found similar variability to be present in their time-resolved spectroscopy. The Lomb-Scargle \citep{Lomb-1976,Scargle-1982} periodograms for both their photometric and spectroscopic data show the strongest peaks at $\sim8.4$ cycles d$^{-1}$, leading to their proposed $P_{\mathrm{orb}}=2.85\pm0.01$h to be the true orbital period. However, it is evident from their periodograms that there are a number of significant peaks close to the chosen frequency, one of which is the previously determined superhump period of 3.24h \citep{Zurita-2008}. It is clear that the timing analysis performed by N14 suffers from aliasing, which is most likely due to the low number of photometric observations obtained, many of which do not cover a complete orbital cycle. \citet{Zurita-2008} obtained 20 photometric observations of $\sim6$h per night, of significantly greater phase coverage. In addition, the variability of the nightly light curves is non-sinusoidal, meaning the phase-dispersion minimisation technique (PDM; \citealt{Stellingwerf-1978}) allowed a more accurate determination of the periodicities present than Lomb-Scargle time-series analysis. We therefore choose $P_{\mathrm{orb}}\approx3.24$h as the orbital period of \source. \\
\indent To further investigate this disagreement over which is the correct value of $P_{\mathrm{orb}}$, we re-examined the light curves presented by \citet{Zurita-2008}. We selected $R$-band photometry from four consecutive nights (27-30 Jun 2007) of observations, in order to compare with the variability demonstrated in the spectroscopy of N14 A subset of the \citet{Zurita-2008} light curves was chosen such that each night of photometry matched the observation duration of the corresponding spectroscopic observation of N14. For example, the 27 Jun 2007 observation was cut from 5.09h to 1.97h to match the length of the Aug 06 2013 spectroscopic observation by N14. We then calculated the PDM and Lomb-Scargle periodograms, separately, for the photometry that matched the sampling of the N14 spectroscopy, and these are presented in Fig. \ref{periodograms}. The upper two panels show that when the data has the same sampling as that of N14, aliasing becomes a serious problem.\\
\indent The Lomb-Scargle periodogram (Fig. \ref{periodograms}; middle panel) shows a very similar structure to N14's Fig. 4, with a number of peaks showing high power, including those corresponding to periods of $\approx3.24$h and $\approx2.85$h. The morphology of the periodograms in Fig. \ref{periodograms} highlights the difficulty of disentangling which is the correct periodicity. The longer timebase and more extensive dataset of \citet{Zurita-2008} removes this problem by including many more complete orbital cycles, with $\approx3.24$h emerging as the most likely value of $P_{\mathrm{orb}}$.\\
\indent This is further illustrated in the bottom panel of Fig. \ref{periodograms}, which shows the Lomb-Scargle periodograms of two simulated light curves. These have the same filtering applied to them as discussed above (i.e. the same sampling and duration as the N14 spectroscopy) but are simulated as a perfect sine curve with periodicities of 3.24h and 2.85h. The periodograms of the simulated data are almost indistinguishable from one another, highlighting the problem of aliasing. Indeed, the periodogram of the $P_{\mathrm{orb}}=2.85$h exhibits the highest peak at $3.24$h, making it difficult to identify which is the true periodicity. Only with a longer data-set, such as that of \citet{Zurita-2008}, will the effects of aliasing be reduced and the true value of $P_{\mathrm{orb}}$ become more apparent. This is seen in Fig. \ref{simulated_LS}, which shows the Lomb-Scargle periodograms of two sine curves with periodicities of 3.24h and 2.85h, created using the same sampling as the full light curves (i.e. 20 nights of photometry) presented by \citet{Zurita-2008}. It is clear from Fig. \ref{simulated_LS} that with a longer data-set, though aliasing is still apparent, it is more straightforward to determine which is the true period.\\
\indent Contrary to N14, $P_{\mathrm{orb}}\approx3.24$h puts \source~outside the so-called `period-gap' of binary systems in the range 2.15--3.18h. In the more numerous accreting white dwarf (AWD) binaries, the well-determined period gap \citep{Knigge-2011} is explained by the switching off of the magnetic braking mechanism (which slows down the spin period of the companion star and hence reduces $P_{\mathrm{orb}}$). This causes the donor star to shrink and mass-transfer stops due to the companion not filling its Roche lobe. It is not clear if there is the same period gap in BHXRTs as in AWDs (see e.g. \citealt{King-1996}). Nevertheless, we find that the $P_{\mathrm{orb}}$ of \source~is close to the edge of this gap and thus may well be still filling its Roche lobe, which is evident from the persistent disc emission seen at optical and X-ray wavelengths. However, it is still unclear why the source has been persistent since its discovery as a transient in 2005.\\

\subsection{Companion Star Features}
We do not find the lack of features associated with the companion star to be surprising. \source~has been a persistent X-ray source since its initial outburst in 2005 and is much brighter than the pre-outburst limit of the DSS (R$\lesssim21$), hence the donor must be fainter than this. Furthermore, the optical spectrum is very typical of that of an accretion disc, exhibiting very few features other than the broad, double-peaked He\small II \normalsize4686\AA~and H$_{\alpha}$ emission lines. The apparent companion features present in the optical spectrum of N14 remain unidentified, as (a) they are not at the wavelengths of any normal late-type stellar features and (b) we find no evidence for their presence in our higher resolution WHT spectrum (Fig. \ref{trail}). Furthermore, the lack of any significant correlation with a wide range of late-type stellar template spectra suggests that any companion star features are washed out by the accretion disc. \\
\indent However, it could be possible that the spectral templates we used do not accurately represent the spectral type of the companion star in \source. For this reason we also cross-correlated each spectrum with templates created by averaging all the spectra, apart from the target spectrum of the cross-correlation. However, the CCFs showed none of the signals seen by N14 (Fig. \ref{CCF}). We therefore conclude that there are no spectral features associated with the companion present in our spectra of \source. \\
\indent The noise in our mean spectrum (Fig. \ref{spec}) indicates that we would have been able to detect normal ($\sim$G-K star) absorption features had they been present at the $\sim2\%$ level with respect to the continuum (as with N14's spectrum). Since J1753 had $V\sim17$ at the time of our observations \citep{Neustroev-2015} this implies that the donor must be fainter than $V\sim21$, which is consistent with its non-detection in the DSS as noted earlier.  We again point out that the so-called absorption features detected by N14 do not correspond with those of any normal late-type star and imply a donor that is brighter than this limit.\\
\indent \emph{What could explain the apparently moving features seen by N14?} The Boller and Chivens spectrograph on the 2.1m telescope at the Observatorio Astron\'{o}mico Nacional in Mexico was used to obtain the results presented by N14. They chose a wide slit width of 2.5" despite the reported much better seeing conditions during the time of the observations. We point out that seeing variations and imperfect telescope pointing can lead to variations in the illumination profile of the source on the CCD. In particular, the 2.5$"$ slit corresponds to 65$\mu$m at the spectrograph's plate scale of 38.4$"$ mm$^{-1}$, which represents a wavelength spread of 8.1\AA~with their 400 lines mm$^{-1}$ grating. If the seeing is much better than 2.5$"$, then velocity shifts approaching 600 km s$^{-1}$ could be induced by telescope tracking errors causing a shift of the stellar image within the slit.\\
\indent However, this movement within the slit would only cause the variability seen if the N14 absorption features are real, yet we have been unable to confirm their existence. One explanation for this is to note that, when performing the cross-correlation, N14 appear to mask only the disc emission and night sky lines, which suggests that the interstellar absorption lines such as the DIBs and Na D were included in the calculation of the CCF. Therefore, telescope tracking errors could have given the impression of these features moving and could be responsible for the apparent RV variability that is seen by N14.\\
\indent We also note that Fig 2. of N14 shows that the blue portion of the spectrum exhibits dramatic variability ($\sim2\times$ in flux) over the course of the observations, and this may also be a manifestation of the source drifting within the wide slit. Indeed, this interpretation is strongly supported by the stability of the B-band observations, which have only ever shown very low level variability ($<0.2$mag).\\

\subsection{Mass of the primary}
\label{sec:mass}
Using the recently derived FWHM(H$_{\alpha}$)-$K_2$ correlation \citep{Casares-2015}, we can obtain an independent estimate of the likely mass, $M_1$, of the primary. We find, conservatively, $M_1>7.4\pm1.2M_{\odot}$, which is a strong indication that the primary is almost certainly a BH, as also indicated by its spectral properties. N14 claim that it is a low mass BH ($<5M_{\odot}$) based on their companion star radial velocity curve yielding a $K_2$ velocity of $\sim380$ km s$^{-1}$, significantly lower than the estimates we present in this work. However, we are not able to replicate these results with our spectra. Therefore we suggest that the mass of the BH is much higher than previously suggested by N14 and hence there is no spectroscopic evidence for a low-mass BH in \source. \\
\indent N14 claim that \source~lies in the so-called `mass gap,' a gap in the distribution of known compact object masses in the range $2-5M_{\odot}$ \citep{Bailyn-1998,Ozel-2010}. The presence of sources in the mass gap may give an indication of the formation scenario of such systems, with BH masses in the range $2-5M_{\odot}$ suggestive of a delayed supernova explosion scenario \citep{Belczynski-2012}. However, with the mass estimate inferred from the H$_{\alpha}$ FWHM we instead place \source~well outside the mass gap, instead falling in the canonical BH mass distribution around $\sim7M_{\odot}$ \citep{Ozel-2010}.

\section{Conclusions}
\source~has now been active for >10 years, which has made it difficult to calculate the system's parameters using the normal methods applied to LMXB BH systems. We have obtained and analysed medium resolution spectroscopy in order to compare our results to those of N14, who claim a very low-mass primary and a short (2.85h) orbital period. We find that we cannot replicate any of the results of N14, finding no spectroscopic evidence of the companion star. Instead, we derive a much higher compact object mass ($M_1>7.4\pm1.2M_{\odot}$), leaving no doubt that the source contains a BH. We also find that the shorter $P_{\mathrm{orb}}=2.85$h preferred by N14 is likely a result of aliasing, and our analysis supports a $P_{\mathrm{orb}}$ of $3.24$h found by \citet{Zurita-2008}, which is more accurately interpreted as the (very fractionally longer) superhump period.

\section*{Acknowledgements}
The authors would like to thank the anonymous referee for comments and suggestions which have helped improve this manuscript. AWS would like to thank Louise Wang for useful discussions regarding Doppler Tomography. JC acknowledges support by DGI of the Spanish Ministerio de Educaci\'{o}n, Cultura y Deporte under grants AYA2013-42627 and  PR2015-00397. Also to the Leverhulme Trust through grant VP2-2015-04. JHVS acknowledges financial support from CONACYT (Mexico) and the University of Southampton. This work made use of Tom Marsh's \textsc{molly} and \textsc{doppler} software packages. The authors would also like to thank Cristina Zurita Espinosa for providing us with her 2007 photometry of Swift J1753.5-0127.




\bibliographystyle{mnras}
\bibliography{references_REV1.bib} 








\bsp	
\label{lastpage}
\end{document}